\documentclass{appolb}
\usepackage{graphicx}
\usepackage[utf8]{inputenc}

\begin{document}
\title{Holograpic Glueball Decay%
\thanks{Based on a talk by F.\ Brünner at Excited QCD 2014, Bjelasnica
Mountain, Sarajevo}%
}
\author{Frederic Brünner, Denis Parganlija and Anton Rebhan
\address{Institut für Theoretische Physik, Technische Universität Wien, \\Wiedner Hauptstraße 8-10, 1040 Vienna, Austria}
}

\maketitle
\begin{abstract}
We announce new results on glueball decay rates 
in the Sakai-Sugimoto model, a realization of
holographic QCD from first principles that has only one coupling
constant and an overall mass scale as free parameters. 
We extend a previous investigation by Hashimoto, Tan, and Terashima
who have considered the lowest scalar glueball which
arises from a somewhat exotic polarization of supergravity modes
and whose mass is uncomfortably small in comparison with lattice results.
On the other hand, the scalar glueball dual to the dilaton turns out
to have a mass of about twice the mass of the rho meson (1487 MeV),
very close to the scalar meson $f_0(1500)$ that is
frequently interpreted as predominantly glue. Calculating the
decay rate into two pions 
we find a surprisingly good agreement with 
experimental data for the $f_0(1500)$.
We have also obtained decay widths for tensor and excited scalar glueballs,
indicating universal narrowness.
\end{abstract}
  
\section{Holographic Glueballs}

Glueballs, color-neutral bound states of gluons, 
are a theoretical prediction of QCD with reliable
quantitative results on the mass 
spectrum available from lattice simulations,
in particular for
pure Yang-Mills theory \cite{Morningstar:1999rf,Chen:2005mg}.
Experimental 
evidence for the existence of glueballs is, however, still elusive
\cite{Klempt:2007cp,Ochs:2013gi},
because glueballs are expected to mix with isoscalar
quark-antiquark bound states
that have identical quantum numbers \cite{Lee:1999kv}.
In order to identify the lightest (scalar) glueball,
whose mass is expected to lie between 1 GeV and 2 GeV, 
additional information such as decay rates (as opposed
to decay constants) would
be needed from first-principles approaches, but
traditional methods like lattice QCD or chiral perturbation
theory lack the ability to perform such calculations. 

A new, promising direction for dealing with strongly coupled
gauge theories and thus a possible avenue to
a theoretical description of glueballs is given 
by so-called holographic QCD, which is not restricted to
Euclidean spacetime signature.
It is based on (or, in phenomenological bottom-up models, motivated by)
the AdS/CFT correspondence, a conjectured exact
duality between certain
four-dimensional quantum field theories and higher-dimensional superstring theories \cite{Maldacena:1997re,Aharony:1999ti}, which in the limit
of infinite number of colors and 't Hooft-coupling
become accessible in the form of weakly coupled supergravity.

Here we shall be interested in so-called top-down models that
have a well-defined superstring theory as their basis and which
correspondingly have minimal freedom to adjust parameters.
A top-down model of low-energy nonsupersymmetric QCD was
pioneered by Witten \cite{Witten:1998zw}, who has
formulated a holographic dual of a five-dimensional super-Yang-Mills
theory compactified on a circle of radius $R=1/M_{\rm KK}$, 
which breaks supersymmetry and leaves only gauge bosons
as low-energy degrees of freedom. In the limit of infinite Kaluza-Klein
mass $M_{\rm KK}$ one obtains pure Yang-Mills theory
in four dimensions, however
calculations using the supergravity approximation are
only possible at finite $M_{\rm KK}$ and large coupling.
This theory allows for a dual description of glueballs, which correspond holographically to fluctuations in the background geometry \cite{Csaki:1998qr,Constable:1999gb}. A full analysis of 
these modes was given in \cite{Brower:2000rp} and their spectrum was derived, 
which bears a striking resemblance to the spectrum
obtained in lattice gauge theory
when one of the glueball states, e.g.\ the lowest tensor glueball,
is matched to the lattice result.

\section{Interaction with chiral quarks}

A possibility to add chiral quarks to the Witten model
described above was found by Sakai and Sugimoto \cite{Sakai:2004cn,Sakai:2005yt}
in the form of pairs of flavor D8-$\overline{\mbox{D8}}$-branes
intersecting the compactification circle at antipodal points.
In the Sakai-Sugimoto model chiral symmetry breaking (as well
as symmetry restoration at high temperature) has
a simple geometric interpretation. It features 
massless pions and massive vector and axial-vector mesons
corresponding to gauge fields on the D8-branes
described by a Dirac-Born-Infeld action, which, after integrating out a 
four-dimensional sphere and expanding to quadratic order in the field strength, is given by
\begin{equation}
 S_{\rm DBI}=
\frac{g_{\mathrm{YM}}^2 N_c^2}{216\pi^3}
\int d^4x\, dz\, \textrm{Tr}\left[\frac{1}{2}K^{-1/3}F^2_{\mu\nu}+K F^2_{\mu z}\right].
\end{equation}
Here $z$ is the holographic radial coordinate with $K=1+z^2$,  
$g_{\mathrm{YM}}$ is the 
four-dimensional Yang-Mills coupling 
at the scale $M_{\rm KK}$, and $N_c$
the number of colors. 
The pions and vector mesons are contained in the nonabelian flavor gauge field
on the D8-branes, $A_\mu=\psi_1(z)\rho_\mu(x^\nu)$ and $A_z=\phi_0(z)\pi(x^\nu)$,
where $\phi_0(z)\propto 1/K$, and $\psi_1(z)$ is the lowest fluctuation
mode along the holographic direction.
Matching the mass of the rho meson and the pion decay constant $f_\pi$
to their experimental values fixes the only free parameters $g_{\mathrm{YM}}$
and $M_{\rm KK}$ to\footnote{In the original and published
version of Ref.~\cite{Sakai:2004cn,Sakai:2005yt} an error in the normalization of
the D8-brane action for the multi-flavor case led to a different
value of $\lambda \simeq 8.3$ that was corrected later in the
e-print version. Unfortunately, this error has not been corrected
in the paper by Hashimoto, Tan and Terashima \cite{Hashimoto:2007ze}
on glueball decay. Corrected results for the decay rates discussed
therein will be presented in Ref.~\cite{BPR}.\label{footnote1}}
\begin{equation}
\lambda\equiv g_{\mathrm{YM}}^2 N_c \simeq 16.63,\qquad
M_{\rm KK} \simeq 949\,{\rm MeV}
\end{equation}
with $N_c=3$.
Doing so reproduces quite nicely the observed
spectrum of the next heavier vector and axial-vector mesons in QCD
(while recent lattice simulations and extrapolations to the chiral
limit and 
large-$N_c$ show stronger 
deviations \cite{Bali:2013kia}).

The effective Lagrangian for the mesons then also allows one to calculate
the decay rate of a rho meson into two pions with the result
\begin{equation}\label{eq:rhowidth}
 \frac{\Gamma_{\rho\rightarrow\pi\pi}}{m_\rho}=\frac{7.659}{g_{\rm YM}^2 N_c^2}
=0.1535
\end{equation}
which is remarkably close to the experimental value $\Gamma_{\rho\rightarrow\pi\pi}/m_\rho\approx 0.19$ \cite{Beringer:1900zz}. (This decay rate was calculated already in \cite{Hashimoto:2007ze}, but their final numerical result differs from ours by a factor of $2$ due to the error mentioned in footnote~\protect\ref{footnote1}.)

This result is quite encouraging to also calculate the decay rates of
glueballs into pions and other mesons within the Sakai-Sugimoto model.
This was already carried out by Hashimoto, Tan and Terashima \cite{Hashimoto:2007ze} for the lowest lying glueball in the spectrum obtained
in \cite{Brower:2000rp} which was compared with
the $f_0(1500)$ isoscalar meson that is frequently interpreted
as being predominantly glue \cite{Amsler:1995td,Close:2005vf,Janowski:2011gt}
(for alternative scenarios see e.g.\ 
\cite{Klempt:2007cp,Ochs:2013gi,Albaladejo:2008qa}).

However, with the overall mass scale
being fixed by the Sakai-Sugimoto model after matching 
the mass of the rho meson,
the lowest scalar glueball mode turns out to be much too light
for this identification -- at 855 MeV
it is only 10\% heavier than the rho meson.\footnote{Modifications
of the Sakai-Sugimoto model corresponding to a non-maximal separation
of D8 and $\overline{\textrm{D8}}$ branes would not help, but
only aggravate this problem.}
Of course, the Sakai-Sugimoto model can a priori not be expected
to be quantitatively accurate to any degree, but other mass ratios
typically come out much better.
The resolution of this discrepancy may well be in corrections
beyond the leading terms of the supergravity approximation. However,
the next scalar mode in the supergravity spectrum would fit almost
perfectly given its mass $M\approx
1.567 M_{\rm KK}\approx 1487$ MeV
(bearing in mind that various $f_0$ states below 2 GeV 
are considered as lowest scalar glueball candidates). This scalar glueball
is coming from
the lowest dilaton mode, 
which in bottom-up AdS/QCD models usually corresponds to the lowest-lying glueball
\cite{BoschiFilho:2002ta,Colangelo:2007pt,Forkel:2007ru}.
In fact, in the context of the analogous problem
in QCD$_3$ obtained from a supersymmetry-breaking
circle compactification of AdS$_5\times S^5$,
it has been argued that the lowest mode in the spectrum obtained
in \cite{Brower:2000rp}, which is associated to an
``exotic'' \cite{Constable:1999gb} polarization of gravitational modes,
should be discarded after all \cite{Terning:2006bq}.
In the following we shall therefore consider the dilatonic holographic glueball
in the Sakai-Sugimoto model
and calculate its decay rate.

The ten-dimensional action for the dilaton $\Phi$ corresponding to a scalar glueball is given by 
\begin{equation}
 S_{\mathrm{Dil}}
\propto\int d^{10}x\sqrt{-g}e^{-2\Phi}\left[R(g_{ab})+4\partial_\mu\Phi\partial^\mu\Phi\right],
\end{equation}
\noindent where $g_{ab}$ is the metric of the Witten model and $R(g_{ab})$ the Ricci scalar. 
The equations of motion corresponding to
this action allow one to determine the spectrum. (This was carried out 
in \cite{Brower:2000rp} in an 11-dimensional supergravity setting, from
which the dilaton mode follows upon dimensional reduction.) 

Combining the dilaton fluctuations about
the (nontrivial) dilaton
background, $\Phi=\Phi_0(z)+\phi$, with the DBI action for pions and mesons, we arrive at an effective four-dimensional interaction Lagrangian. If we assume the dilaton to be of the form
$\phi=H(z)G(x^\nu)$, after integrating out the holographic coordinate the coupling of a single scalar glueball to two pions is given by 
\begin{equation}
 S_{G\pi\pi}=\mathrm{Tr}\int d^4x\frac{1}{2}\tilde c_1\partial_\mu\pi\partial^\mu\pi G,
\end{equation}
where the constant $\tilde c_1$ can be numerically determined as 
\begin{equation}
\tilde c_1=\frac{14.92}{g_\mathrm{YM}N_c^{3/2}M_{\rm KK}} 
\end{equation}
for the lightest dilaton mode (full details will be given in \cite{BPR}).

With these ingredients, it is straightforward to determine the corresponding decay rate. It is given by 
\begin{equation}
 \frac{\Gamma_{G\rightarrow\pi\pi}}{M}=\frac{3|\tilde c_1|^2 M^2}{128\pi}=\frac{4.076}{g_{\rm YM}^2 N_c^3}
=0.027.
\end{equation}
Note that compared to the decay rate of the rho meson evaluated above this is
parametrically suppressed by an additional factor $1/N_c$. 

\goodbreak
The experimental value for the decay of the $f_0(1500)$ isoscalar into two pions has been determined as $\Gamma_{f_0(1500)\rightarrow\pi\pi}/M_{f_0(1500)}=0.025(3)$ 
\cite{Beringer:1900zz}, which matches the holographic result
surprisingly well. 
This nearly perfect agreement is of course somewhat fortuitous.
Let us consider one alternative to fixing the coupling constant in
the Sakai-Sugimoto model: instead of matching the pion decay
constant $f_\pi$ we could also choose to match the
rho meson width obtained in Eq.~(\ref{eq:rhowidth}),
leading to a slightly higher width for the dilatonic glueball,
to wit, ${\Gamma_{G\rightarrow\pi\pi}}/{M}=0.034$.
This is still sufficiently close to the observed width
of the $f_0(1500)$ to be taken as very interesting.
It certainly does encourage deeper investigations
of glueball properties in the almost parameter-free Sakai-Sugimoto model.

We will further elaborate on glueball decay rates 
in the Sakai-Sugimoto model in \cite{BPR}, 
discussing systematically all of the lowest-lying glueballs
and also the decay into four pions.
As an apparently generic feature we have found narrow widths
also for the tensor glueball and the excited scalar glueballs
arising from dilaton modes.

\bigskip

F.B.\ and D.P.\ were supported by the Austrian Science Fund FWF,
project no.\ P26366.

\newpage

\section*{Erratum}

\setcounter{equation}{4}
The effective Lagrange density for the coupling of the (predominantly) dilatonic glueball
given in Eq.~(5) is incomplete, because it did not take into account
all metric perturbations that are induced by the dilaton perturbation $\phi$.
Eq.~(5) should instead read
\begin{equation}
 S_{G\pi\pi}=\mathrm{Tr}\int d^4x\frac{1}{2}\tilde c_1\partial_\mu\pi\partial_\nu\pi 
\left(\eta^{\mu\nu}-\frac{\partial^\mu \partial^\nu}{M^2}\right)G,
\end{equation}
with $\tilde c_1$ now given by
\begin{equation}
\tilde c_1=\frac{17.23}{g_\mathrm{YM}N_c^{3/2}M_{\rm KK}}.
\end{equation}
The result for the decay width 
changes to
\begin{equation}
 \frac{\Gamma_{G\rightarrow\pi\pi}}{M}=\frac{3|\tilde c_1|^2 M^2}{512\pi}=\frac{1.359}{g_{\rm YM}^2 N_c^3}
=0.009.
\end{equation}
Alternatively fixing the 't Hooft coupling such that the rho meson width is
matched to experimental data gives ${\Gamma_{G\rightarrow\pi\pi}}/{M}=0.011$.
Both results are thus below 
the experimental value of 0.025(3) for the decay of the $f_0(1500)$ isoscalar into two pions.
Given that the larger decay width of the $f_0(1500)$ may be due to admixture of 
less narrow $q\bar q$ states, 
we still take this as an encouraging result
of our attempt to describe glueball properties
through the almost parameter-free Sakai-Sugimoto model.\footnote{In
this context we note that the recent paper by S.~Janowski, F.~Giacosa, and D.~Rischke,
arXiv:1408.4921, 
which extends Ref.~\cite{Janowski:2011gt} to the three-flavor case,
no longer identifies the $f_0(1500)$ isoscalar as a predominantly glueball state,
but instead favors the $f_0(1710)$, which has a smaller decay width into two pions.}

\end{document}